\documentclass[conference]{IEEEtran}
\IEEEoverridecommandlockouts

\usepackage{cite}
\usepackage{amsmath,amssymb,amsfonts}
\usepackage{xspace}
\usepackage{tabularx}
\usepackage{colortbl}
\usepackage{subfig}
\usepackage{multirow}
\usepackage{booktabs}
\usepackage{listings}
\usepackage{arydshln}
\usepackage{balance}
\usepackage{xcolor}
\usepackage[hidelinks]{hyperref}
\usepackage[colorinlistoftodos,prependcaption,textsize=tiny]{todonotes}
\usepackage[capitalise]{cleveref}

\newcommand*{\belowrulesepcolor}[1]{%
  \noalign{%
    \kern-\belowrulesep
    \begingroup
      \color{#1}%
      \hrule height\belowrulesep
    \endgroup
  }%
}
\newcommand*{\aboverulesepcolor}[1]{%
  \noalign{%
    \begingroup
      \color{#1}%
      \hrule height\aboverulesep
    \endgroup
    \kern-\aboverulesep
  }%
}

\usepackage[colorinlistoftodos,prependcaption,textsize=tiny]{todonotes}

\usepackage{fbox}
\newcommand{\rqsummary}[2]{
        \vspace{2mm}
        \noindent
        \fbox{%
            \parbox{.97\linewidth}{%
                    \textbf{#1 Summary.}
                #2
            }%
        }%
        \vspace{2mm}
}%

\newlength\MAX  

\newcommand{\DrawPercentageBar}[1]{%
  \begin{tikzpicture}
    \fill[color=black]   (0.0 , 0.0) rectangle (#1*5ex , 1.7ex );
    \fill[color=lightgray] (#1*5ex  , 0.0) rectangle (5.0ex, 1.7ex);
  \end{tikzpicture}%
}

\def\BibTeX{{\rm B\kern-.05em{\sc i\kern-.025em b}\kern-.08em
    T\kern-.1667em\lower.7ex\hbox{E}\kern-.125emX}}

\begin{document}

\title{{Belonging Beyond Code: Queer Software Engineering and Humanities Student Experiences}
}

\author{\IEEEauthorblockN{Emily Vorderwülbeke}
\IEEEauthorblockA{\textit{University of Passau} \\
Passau, Germany \\
vorderwuel@fim.uni-passau.de}
\and
\IEEEauthorblockN{Isabella Graßl}
\IEEEauthorblockA{\textit{University of Passau} \\
Passau, Germany \\
isabella.grassl@uni-passau.de}
}

\maketitle

\begin{abstract}
Queer students often encounter discrimination and a lack of belonging in their academic environments. This may be especially true in heteronormative male-dominated fields like software engineering, which already faces a \emph{diversity crisis}. 
In contrast, disciplines like humanities have a higher proportion of queer students, suggesting a more diverse academic culture.
While prior research has explored queer students' challenges in STEM fields, limited attention has been given to how experiences differ between the sociotechnical, yet highly heteronormative, field of software engineering and the socioculturally inclusive humanities.
This study addresses that gap by comparing 165 queer software engineering and 119 queer humanities students experiences. 
Our findings reveal that queer students in software engineering are less likely to be open about their sexuality, report a significantly lower sense of belonging, and encounter more academic challenges compared to their peers in the humanities. Despite these challenges, queer software engineering students show greater determination to continue their studies.
These insights suggest that software engineering could enhance inclusivity by adopting practices commonly seen in the humanities, such as integrating inclusive policies in classrooms, to create a more welcoming environment where queer students can thrive.
\end{abstract}

\begin{IEEEkeywords}
LGBTQ+, queer students, software engineering.
\end{IEEEkeywords}

\section*{Prelude}
Picture a conversation among software engineering students, prompted by the emerging topic of diversity in software engineering~\cite{albusays2021d,de2023lgbtqia}, reflecting on their experiences in their studies.
The male heterosexual students assert, \emph{“Sexuality is never a topic as---in my academic experience---nobody actually cares,” (P201)}\footnote{\emph{P} followed by a number represents a participant ID from our study.} and, \emph{“It’s delusional to think your sexuality hinders you in your studies; get over it.” (P263)}. For them, sexual identity seems irrelevant and unrelated to academic success. 

In contrast, the \emph{queer}\footnote{We use \emph{queer} as an inclusive umbrella term to encompass all sexual identities of individuals who do not identify as heterosexual~\cite{gonzalez2024}.} students share a different reality: \emph{“I have faced negative and exclusionary behavior, including being ignored or shunned by peers and faculty, which has hindered my ability to concentrate on my studies and affected my well-being” (P767)}, and \emph{“some of my interactions are not genuine since I am hiding an essential part of who I am. This had a negative impact on my lecture attendance and classroom participation.” (P760)}. 
Their words highlight the emotional and social burdens of navigating their studies while hiding parts of their (sexual) identity.

\section{Introduction}
\label{sec:introduction} 
Sexuality is rarely discussed in academic environments, as many believe it does not matter in professional and educational settings~\cite{cech2013,cech2013a}. However, this perspective often only applies for cisgender heterosexual male individuals~\cite{miller2021,canedo2023}. 
In contrast, queer individuals face significant challenges, which begin as early as their educational experiences, particularly within \emph{Science, Technology, Engineering and Mathematics} (STEM) fields~\cite{hughes2018coming,jennings2020}. 
In these fields, queer individuals are underrepresented \cite{casper2022revealing} and exhibit higher dropout rates, compared to their heterosexual peers \cite{hughes2018coming}. This disparity can be attributed to exclusion, disadvantages, and discrimination against queer individuals \cite{cech2018lgbtq,cech2009} as well as a low \emph{sense of belonging}~\cite{hagerty1992}, which has been shown to correlate strongly with continued enrolment in these studies \cite{stout2016lesbian}. 

At the core of these experiences lies the fundamental human desire to feel welcomed and appreciated for one's authentic self~\cite{maslow1943}. 
While some students are able to pursue their studies without encountering  obstacles beyond the academic demands of the discipline~\cite{canedo2023}, others must constantly manage their identity in an environment that remains far from neutral~\cite{kosciw2015}. 
Assessing whether students feel safe expressing their identities and understanding the impact on their well-being is crucial~\cite{johnson2022,strayhorn2018}. 
While prior research has predominantly focused on STEM fields broadly, it is essential to explore these issues within specific disciplines like software engineering which involves \emph{sociotechnical} dimensions by centring the human element~\cite{carver2021}. Notably, this field has shown particularly negative effects for queer professionals within the industry~\cite{desouzasantos2023}.

In software engineering, diverse teams are vital to improving creativity, quality~\cite{pieterse2006software, crawford2012agile}, and addressing varied customer needs~\cite{mannix2005differences}, where effective teamwork depends not just on technical skills but also on social dynamics~\cite{mannix2005differences, pieterse2006software, crawford2012agile}. 
However, diversity is often underexamined in software engineering, which has historically marginalized queer individuals.

This marginalization is less pronounced in disciplines like the \emph{humanities} including studies such as linguistics, history or media where the focus on social issues~\cite{greathouse2018queer, stout2016lesbian} has cultivated more inclusive environments~\cite{roseth2019features}. Unlike software engineering, where human-centered concerns are still secondary to technical priorities, humanities disciplines actively engage with social identities and foster spaces where students may feel more comfortable being open about their identities~\cite{linley2018}.

In this paper, we compare the experiences of 236 queer students with 165 in software-related studies and 119 in humanities-related studies, along with 48 non-queer students. Including non-queer students provides anecdotal insights to highlight the unique challenges faced by queer students.

To contextualize these experiences, we need to examine the level of identity visibility, i.e., the \emph{outness}~\cite{kosciw2015}, among queer students. Visibility plays a critical role in shaping their experiences in educational settings, especially when comparing traditionally technical fields like software engineering with more socially engaged disciplines like the humanities~\cite{linley2018}. 
Outness interacts with another key factor---the \emph{campus climate}~\cite{garvey2015influence}---which reflects the overall atmosphere toward queer individuals in their academic settings. By examining outness and campus climate, we can better understand how these factors impact queer students’ experiences across these two disciplines.


Accordingly, our first research question (RQ) aims to investigate the level of outness and the experiences among students:
\textit{\textbf{RQ1:} How do levels of outness and campus climate compare for queer students in software engineering and humanities?}

Feeling a sense of belonging is crucial since it influences students' mental health and academic success~\cite{strayhorn2018,hagerty1992,baumeister2017}. 
This feeling of being included or excluded might vary between queer students across different disciplines~\cite{stout2016lesbian}, particularly given the traditionally masculine and heteronormative culture in fields like software engineering~\cite{miller2021,hughes2017}.

\textit{\textbf{RQ2:} How does the sense of belonging compare between queer students in software engineering and humanities?}

According to prior studies, students who feel marginalized, especially in the queer community, are more likely to leave their studies~\cite{stout2016lesbian, mooney2021investigating, de2023lgbtqia}, making it important to investigate whether queer students in software engineering and humanities are dropping out at different rates---and \emph{why}. This allows us to assess whether certain environments are more likely to pushing students out of their academic careers.

\textit{\textbf{RQ3:} How do dropout rates compare between queer students in software engineering and those in humanities?}

By comparing the experiences of queer students across multiple fields, we aim to highlight existing issues while also propose strategies derived from humanities that could be implemented to enhance the experiences of queer students. 

In particular, our findings indicate that software engineering students exhibit a significantly lower sense of belonging, particularly in terms of feeling welcome and respected by their peers, and are less likely to be out with faculty and fellow students. In contrast, while queer students in the humanities report more positive experiences--—partially due to higher levels of outness---they exhibit a higher dropout rate compared to their computer science counterparts. Overall, queer students, regardless of their field, are more likely to leave their studies compared to their non-queer peers. 
Thus, this paper gives actionable insights to create a more supportive environment for queer students, which can also inform practices in professional software development environments.

\section{Background \& Related Work}\label{sec:background}
\emph{Queer} refers to individuals who do not conform to heteronormative  sexuality as well as those whose gender expressions, performances, and understandings challenge or resist traditional gender binaries~\cite{jagose1996}. While the term \emph{queer} is intentionally fluid and open-ended, in this study, we limit its use to individuals who do not identify as heterosexual~\cite{gonzalez2024}.

\subsection{Queerness and its Impact}
Diversity, including sexual orientation as its internal dimension, is integral to individuals' identities, much like age, gender, and ethnicity~\cite{gardenswartz2003}. The sexual orientation has diverse facets where being \emph{queer}, or belonging to the \emph{LGBTQIA+} community\footnote{LGBTQIA+ is an umbrella term and stands forlesbian, gay, bisexual, transgender/transsexual, queer/questioning, intersex, asexual, where the \emph{+} includes other orientations and identities~\cite{jennings2020}. We use this term only when referring to specific bodies of work where authors referred to these conceptualizations.}, means falling outside the sociocultural established traditional definitions of heterosexual~\cite{jagose1996}.

A central concept within the queer experience is \emph{outness}, which refers to how openly someone shares their sexual orientation or gender identity. While higher levels of outness can foster a stronger sense of authenticity, it can also expose individuals to increased discrimination and hostility, shaped by social and cultural contexts. Conversely, remaining less open can provide protection, but may result in feelings of isolation and emotional burden of hiding one's true self.

In academic settings, a strong \emph{sense of belonging} and a positive \emph{campus climate} are essential for queer students. When students feel included and respected, it boosts their mental health and academic success~\cite{garvey2018impact}. 
Campus climate, reflecting the attitudes and behaviours in the academic environment, heavily influences queer students’ experiences~\cite{greathouse2018queer}. Students who report a higher comfort or warmer perceptions with positive campus climate had more academic success than their peers \cite{garvey2018impact}. However, this climate is influenced by classroom-climate, students gender-conformity and outness, where a more positive classroom climate is correlated with gender conforming students \cite{garvey2015making} and a more negative campus climate is associated with a higher level of outness \cite{brinkworth2016chilly, garvey2015influence, garvey2019queer}.
However, despite strides in equality, queer individuals still face significant discrimination on many campuses, particularly in traditionally heteronormative fields~\cite{ellis2009diversity}.

\subsection{Queerness in STEM: The Case of Computer Science}
STEM disciplines are often characterized by a \emph{dude culture}\cite{miller2021}, where a mentality of \emph{don’t ask, don’t tell}\cite{bilimoria2009} persists. 
This culture not only marginalizes queer individuals but also forces everyone not conforming to the white male heteronormativity to hide their identities~\cite{haverkamp2018,bilimoria2009,leyva2017,leyva2022}. The limited representation of queer individuals in STEM diversity initiatives further exacerbates their marginalization\cite{casper2022revealing}. 

Furthermore, limited research exist on queerness in computer science, especially within higher education~\cite{jennings2020,gonzalez2024}, though studies suggest that LGBTQ+ students face greater health issues and marginalization than their heterosexual peers, restricting academic success and fostering feelings of isolation~\cite{cech2018lgbtq,cech2011navigating}. In physics and other STEM fields, many LGBTQ+ students describe their environment as \emph{chilly}~\cite{brinkworth2016chilly} or uncomfortable~\cite{atherton2016physics}.
As a result, many are forced to adopt \textit{coping strategies} such as \emph{passing} as heterosexual~\cite{cech2011navigating,cech2009}. 
%
A significant portion of queer students in STEM report lower persistence, with some studies finding them more likely to leave their programs compared to their heterosexual peers~\cite{hughes2018coming, casper2022revealing}. In particular, studies of LGBTQ graduate and undergraduate students, show that they are thinking about leaving their degree due to a low \textit{sense of belonging} to the computing community more frequently than their heterosexual peers \cite{stout2016lesbian}, which also was the result of other studies \cite{mooney2021investigating, de2023lgbtqia}. 
This also means that universities are not perceived as \emph{safe spaces} for queer students, where they can be open about their sexual identity~\cite{ellis2009diversity}.



\subsection{Disparities Across Disciplines: The Case of Humanities}
While STEM fields are marked by exclusionary environments for queer students, the humanities appear to offer more supportive spaces~\cite{linley2018}. 
Studies have found that a higher percentage of queer-identifying students are present in fields such as the humanities (20.4\%) \& arts and social sciences (17\%) compared to STEM disciplines(6.7\%)~\cite{greathouse2018queer}. Studies on music and arts departments, in particular, show that students in these areas experience better well-being compared to non-art disciplines~\cite{roseth2019features}. These students are more likely to engage in discussions about their identities, and might benefit from a more inclusive and affirming campus climate~\cite{stout2016lesbian,linley2018}.

While the humanities appear to be more welcoming, there is limited research on the specific experiences of queer students across different academic disciplines. Understanding these differences could help educators create more inclusive environments tailored to various academic fields~\cite{stout2016lesbian}.

\subsection{Invisibility in Software Engineering Research}
Despite growing awareness of the challenges faced by diverse individuals in software engineering~\cite{albusays2021d}, their experiences remain largely invisible in the  research community~\cite{de2023lgbtqia,jennings2020}.

One significant issue is the persistent absence of sexual identity demographics in most quantitative studies, rendering these individuals invisible, particularly to university administrators \cite{greathouse2018queer,casper2022revealing}. For instance, less than 2\% of research articles in higher education journals mention sexual identity demographics~\cite{greathouse2018queer}, and only 3\% of diversity studies in computer science and software engineering education explicitly mention LGBTQ+ students~\cite{de2023lgbtqia}. This lack of attention further marginalizes queer individuals, leaving their specific challenges under-explored. Moreover, the majority of research focuses on queer students from US, UK and Ireland \cite{de2023lgbtqia,ellis2009diversity}. This leads to a \emph{LGB Monolith}, largely focusing on the experiences of white, cisgender, middle-class, and homosexual men and women~\cite{jennings2020}. 

This study aims to bridge this gap~\cite{stout2016lesbian, mooney2021investigating} by comparing the experiences of queer students in  software engineering and humanities to assess the extent to which different academic environments support or undermine the experiences of queer students.

\section{Method}\label{sec:methodology}
This study employs a mixed-methods approach~\cite{creswell1999}, inviting software and humanities students to share their academic experiences in terms of their sexuality.

\subsection{Instrumentation}
We designed the questionnaire based on established literature~\cite{garvey2019queer,mooney2021investigating,senseofbelongingScale2020,stout2016lesbian,tang2023impact}.
To ensure its effectiveness, we conducted a pilot study with five researchers who are actively engaged in software engineering and software engineering education at the \emph{University of Passau (Germany)}.
This pilot study allowed us to evaluate the wording, comprehension, and estimated completion time of the questionnaire. After addressing minor wording issues identified during the pilot, the final questionnaire included a series of items~(\Cref{tab:questionnairestudents}).

Prior to the questionnaire, we explained the purpose and process of the study, including a trigger warning since the survey contains questions about sexuality, which may be sensitive for some participants. Respondents were informed they could stop the survey at any time if they felt uncomfortable, and that all responses would be completely anonymized.


\begin{table}[tb]
\centering
\caption{Questionnaire for all students.}
\label{tab:questionnairestudents}
\resizebox{\columnwidth}{!}{%
\begin{tabular}{lp{6.5cm}l}
\toprule
Var. & Question & Source\\

\midrule
\belowrulesepcolor{gray!25}
\rowcolor{gray!25}\multicolumn{3}{l}{Level of Outness \& Campus Climate (RQ1)} \\
\aboverulesepcolor{gray!25}
SE01	& How \emph{out} are you with professors, faculty, and instructors? & \cite{garvey2019queer}\\
SE02	& How \emph{out} are you with student peers? & [new] \\

CC01	& Have you \emph{personally} experienced any exclusionary (e.g., shunned, ignored), intimidating, offensive and/or hostile conduct (harassing behavior) that has interfered with your ability to learn or your wellbeing in your current field of study because of your sexual orientation or identity? & \cite{atherton2016physics} \\
CC02	& Please briefly explain your answer to the prior question. & \cite{atherton2016physics} \\
CC03	& Have you \emph{personally} experienced any inclusive (e.g., welcomed, acknowledged), encouraging, pleasant and/or friendly conduct (supportive behavior) that has positively impacted your ability to learn or your wellbeing in your current field of study because of your sexual orientation or identity? & [new] \\
CC04	& Please briefly explain your answer to the prior question.& [new] \\

\midrule
\belowrulesepcolor{gray!25}
\rowcolor{gray!25}\multicolumn{3}{l}{Sense of Belonging (RQ2)} \\
\aboverulesepcolor{gray!25}
SB01	& How much do you interact socially with other students in your studies? & \cite{mooney2021investigating} \\
SB02	& How well do people in your current field of study understand you as a person? & \cite{senseofbelongingScale2020} \\
\midrule
SB03	& How connected do you feel to the university staff in your current field of study? & \cite{senseofbelongingScale2020} \\
SB04	& How welcoming have you found your current field of study to be? & \cite{senseofbelongingScale2020} \\
\midrule
SB05	& How much respect do other students in your current field of study show toward you? & \cite{senseofbelongingScale2020} \\
SB06	& How much respect do members of staff in your current field of study show toward you? & \cite{senseofbelongingScale2020} \\
SB07	& How much do you matter to others in your current field of study? & \cite{senseofbelongingScale2020} \\
\midrule
SB08 & How happy are you with your choice to be a student in your current field of study? & \cite{senseofbelongingScale2020} \\
SB09 & How enriching is your experience in your current field of study? & \cite{senseofbelongingScale2020} \\
SB10 & How 'at home' do you feel in your current field of study? & \cite{senseofbelongingScale2020} \\
\midrule
SB11 & Overall, how much do you feel like you belong in your current field of study? & \cite{senseofbelongingScale2020} \\ 

\midrule
\belowrulesepcolor{gray!25}
\rowcolor{gray!25}\multicolumn{3}{l}{Dropout Rate (RQ3)} \\
\aboverulesepcolor{gray!25}
 DP01 &	During your academic career, have you ever seriously considered leaving your graduate program? & \cite{stout2016lesbian} \\
 DP02 &	Please briefly explain your answer to the prior question.  & \cite{stout2016lesbian} \\

\bottomrule
\end{tabular}%
}
\end{table}

\paragraph{\textbf{Outness \& Campus Climate (RQ1)}}
Asking about outness is essential for understanding how openly students express their sexual identity within academic settings, as it can impact their experiences of inclusion or exclusion. The outness might relate to a positive or negative campus climate, which shape the educational experiences of queer students.
We used a well-established outness scale~\cite{garvey2019queer} to determine students' outness with faculty and staff (\emph{SE01}, \Cref{tab:questionnairestudents}), and extended this question to include outness with their student peers (\emph{SE02}) to explore if peer relationships play a distinct role in their sense of belonging. We posed the same questions to non-queer students to provide a comparative perspective.
The responses for both questions were categorized on a 5-point Likert scale where 1 represents \emph{not at all out} and 5 represents \emph{completely out}.

To assess the experiences, we asked students whether they had personally experienced \emph{exclusion} in terms of shunned or harassing behaviour based on their sexuality (\emph{CC01}, \Cref{tab:questionnairestudents}) \cite{atherton2016physics}. 
The responses were categorized (\emph{yes/no}), followed by an open answer field (\emph{CC02}), providing us insights into the sources of these negative encounters. This focus on \emph{personal} experience ensures that the data reflects the students’ direct encounters rather than perceptions of others.

To foster a more comprehensive understanding of campus climate, we also asked students whether they had experienced \emph{inclusive} experiences (\emph{CC03}, \Cref{tab:questionnairestudents}), using the positive phrasing of the negative conduct (\emph{CC01}). This was also followed by an open answer field (\emph{CC04}), where students could elaborate on these positive experiences.

\paragraph{\textbf{Sense of Belonging (RQ2)}} 
Sense of belonging is a critical aspect of student engagement and success~\cite{baumeister2017}. A strong sense of belonging has been shown to improve academic performance, dropout rates, and overall satisfaction. 
To assess students' sense of belonging, we utilized the 10-item sense of belonging scale developed by \emph{Imperial College London} (\emph{SB02--SB011}, \Cref{tab:questionnairestudents})~\cite{senseofbelongingScale2020}. 
We extended it by a question on their social interaction with peers (\emph{SB01}), since this might be particularly relevant for software students~\cite{mooney2021investigating}. A 5-point Likert scale was used, with higher scores indicating a greater sense of belonging~\cite{tang2023impact}.

\paragraph{\textbf{Dropout Rate (RQ3)}}
Understanding the dropout rate is important because it reflects students' dissatisfaction and disconnection from their academic environment, which may be tied to their experiences of exclusion or a lack of belonging.
We assessed dropout rates by inquiring whether students had ever considered dropping out of their studies (\emph{DP01}) \cite{stout2016lesbian}, with responses categorized as \emph{yes/no}. 
To gain more insights into their reasons for considering dropping out, we followed up with an open-ended question (\emph{DP02}, \Cref{tab:questionnairestudents}).

\subsection{Data Collection} 
Our target group consisted of students in software-related fields (e.g., software engineering, web development) and humanities (e.g., arts, linguistics, media). Since we inquired about their experiences with sexuality in general—--rather than explicitly targeting queer individuals to avoid \emph{outing} someone—--both queer and non-queer students participated, as their insights were also relevant. The data collection took place between May and July 2024.
Given the challenges associated with researching a hidden population~\cite{de2023lgbtqia}, we employed a multipronged sampling strategy to maximize the reach and diversity of our participants~\cite{desouzasantos2023,ellard-gray2015}. 

\begin{itemize}
    \item Convenience Sampling: We directly addressed members of the queer community within the \emph{University of Passau (Germany)} and the \emph{University of Victoria (Canada)}, facilitating access to participants who may have been more willing to share their experiences in a familiar environment.
    \item Snowball Sampling: We encouraged initial respondents to share the survey with their peers, creating a snowball effect that helped reach additional participants. This method proved valuable for tapping into networks that might otherwise remain inaccessible.
    \item Social Media Outreach: To further enhance our participant pool, we leveraged the social media platform \emph{Instagram}, targeting queer communities and our followers. 
    \item Purposive Sampling: We also employed purposive sampling through the use of \emph{Prolific}~\cite{Prolific2023}, a platform known for connecting researchers with diverse participant demographics which was successfully used in prior software engineering studies~\cite{russo2022b,baskararajah2021}. This allowed us to reach a wider audience and ensured representation from a broader spectrum of backgrounds and experiences. The screening questions contained their field of studies (software engineering or humanities), being a member of the queer community, and having an approval rate of 100\% to ensure the best possible accuracy. The average reward per hour was \pounds 11.96 and the median time 15 minutes.
\end{itemize}


\subsection{Participants}
\begin{table}[tb]
	\centering
	\caption{Demographics of students by subgroups.\\ \tiny(queer (Q), hetero (H), humanities (HA), software (SE), \emph{Prolific} (prol.), non-\emph{Prolific} (n-prol.)).}
	\label{tab:demo}
	\resizebox{\columnwidth}{!}{%
\begin{tabular}{llrrrrrrr}
	\toprule
	Category &Subcat. & Q & H & HA & SE & prol. & n-prol. & $\sum$ \\
	\midrule
	Age & 18-24 & 138 & 31 & 74 & 95 & 110 & 59 & 169 \\
	& 25-34 & 80 & 15 & 39 & 56 & 69 & 26 & 95 \\
	& $\geq$35 & 18 & 2 & 6 & 14 & 17 & 3 & 20 \\
	\midrule
	Gender & Female & 124 & 12 & 82 & 54 & 102 & 34 & 136 \\
	& Male & 81 & 36 & 25 & 92 & 70 & 47 & 117 \\
	& Other & 31 & 0 & 12 & 19 & 24 & 7 & 31 \\
	\midrule
	Ethnicity
	& Europe & 166 & 42 & 91 & 117 & 127 & 81 & 208 \\
	& N. Am. & 26 & 2 & 13 & 15 & 27 & 1 & 28 \\
	& Asia & 18 & 2 & 8 & 12 & 16 & 4 & 20 \\
	& Africa & 15 & 2 & 6 & 11 & 15 & 2 & 17 \\
	& S. Am. & 11 & 0 & 1 & 10 & 11 & 0 & 11 \\
	\midrule
	Country
	& Europe & 155 & 44 & 86 & 113 & 116 & 83 & 199 \\
	& N. Am. & 51 & 2 & 20 & 33 & 51 & 2 & 53 \\
	& Africa & 27 & 2 & 12 & 17 & 27 & 2 & 29 \\
	& S. Am. & 2 & 0 & 0 & 2 & 2 & 0 & 2 \\
	& Asia & 1 & 0 & 1 & 0 & 0 & 1 & 1 \\
	\bottomrule
\end{tabular}
}
\end{table}

\begin{table}[tb]
	\centering
	\caption{Demographics of students by studies and sexuality.}
	\label{tab:demo_sexuality}
\begin{tabular}{lrrrrr}
	\toprule
	Sexuality & HA & SE & prol. & n-prol. & $\sum$ \\
	\midrule
	Asexual   & 7 & 9 & 11 & 5 & 16 \\
	Bisexual  & 61 & 77 & 124 & 14 & 138 \\
	Demisexual & 1 & 2 & 3 & 0 & 3 \\
	Heterosexual  & 9 & 39 & 5 & 43 & 48 \\
	Homosexual  & 24 & 16 & 30 & 10 & 40 \\
	Pansexual  & 11 & 20 & 20 & 11 & 31 \\
	Queer& 3 & 2 & 1 & 4 & 5 \\
	Other & 3 & 0 & 2 & 1 & 3 \\
	\midrule
	$\sum$  & 119 & 165 & 196 & 88 & 284 \\
	\bottomrule
\end{tabular}%
\end{table}

Our dataset comprises 284 participants (\Cref{tab:demo}), including 184 undergraduate students and 100 graduate students, predominantly aged 18 to 24, from universities in 27 different countries, mainly in Europe. 
It is important to note that the representation of participants from regions like Asia and South America was limited, which may reflect differences in openness about sexual identity across cultures.
Of these participants, 88 were recruited through university and social media channels, while 196 were sourced via \emph{Prolific}.
In terms of study programs, 165 participants were enrolled in software-related programs, and 119 in humanities. For those studying “computer science" or “IT", we asked if their focus is on software engineering. Humanity programs include fields such as animation, literature, linguistics, history, media studies, and sociology.\footnote{\url{https://figshare.com/s/dc18f812716249d58357.}}
A majority (236)  self-identified as non-hetereosexual (\emph{queer}), while 48 identified as heterosexual~(\Cref{tab:demo_sexuality}).


\subsection{Data Analysis}
We used multiple statistical tests to analyse the effects of sexuality and field of study on the dependent variables: negative and positive campus climate, sense of belonging, and drop-out rate.
Since our sample included various sexualities (\Cref{tab:demo_sexuality}), we consolidated all queer identities into the category \emph{queer} and retained the heterosexual group as a separate category for comparison. However, due to the small size of the heterosexual sample in each study group (SE: 9, HA: 39), we focused on comparing queer software and humanities students, using heterosexual results only as anecdotal references. 

In all our analysis we employed a consistent encoding scheme in which \emph{ yes, queer students,} and \emph{software students} were assigned a value of 1, while \emph{no, non-queer students,} and those in the \emph{humanities} were assigned a value of 0. We also used for all statistical tests a significance level at $\alpha\leq 0.05$.

\paragraph{\textbf{Outness \& Campus Climate (RQ1)}}

The differences in the degree of outness to faculty and staff (\emph{SE01}) as well as student peers (\emph{SE02}), between individuals in humanities and software were measured by the Mann-Whitney U-test~\cite{mann1947}. Vargha-Delaney's Â$_{12}$ was employed to assess effect size~\cite{vargha2000}.

To assess whether students in both fields of study experienced a negative (\emph{CC01}) or positive (\emph{CC03}) campus climate based on their sexuality, we conducted a chi-square test~\cite{delucchi1993} using the binary responses (\emph{yes/no}). This analysis was followed by a logistic regression~\cite{hosmerjr2013} to evaluate the influence of outness (\emph{5-point Likert scale}) on negative campus climate.
We analysed the varying types and levels of negative or positive campus climate reported in the open-ended questions ((\emph{CC02, CC04})) through thematic analysis~\cite{braun2012} with open coding~\cite{miles1994}. Two researchers independently developed initial codes, agreed on a coding scheme, and recoded the responses using the finalized framework. The interrater reliability~\cite{cohen1960b} shows a high agreement ($\kappa=0.82$) between the raters.

\paragraph{\textbf{Sense of Belonging (RQ2)}} 
We used the Mann-Whitney U-test to determine whether a significant difference in sense of belonging (\emph{SB01--11}) exist between software and humanities students. Vargha-Delaney's Â$_{12}$ was used to measure effect size. We used ANOVA~\cite{st1989} to evaluate the effect of sexuality on students' sense of belonging within each study program. Additionally, we analysed the effect of the level of outness on students' sense of belonging utilizing linear regression~\cite{weisberg2005} (\emph{both 5-point Likert-Scale}).

\paragraph{\textbf{Dropout Rate (RQ3)}} 

We analysed the dropout rate (\emph{DP01}, \emph{yes/no}) to determine whether the proportion of students wanting to dropout differs between the two study programs, with the chi-square test. To examine the impact of a student's sexuality on their thoughts about dropping out, we conduct a logistic regression analysis. We perform a logistic regression to determine the effect of the sense of belonging on the dropout rate, comparing the results between the two fields of study. 
For open-ended responses on reasons for considering dropping out (\emph{DP02}), we conducted thematic analysis following the procedure used in \emph{RQ1}, where we established and applied a coding scheme. The interrater reliability shows a high agreement ($\kappa=0.86$) between the raters.

\subsection{Ethics}
This study was conducted in accordance with the ethical guidelines of the \emph{University of Passau (Germany)} as well as general standards for human research, particularly vulnerable groups~\cite{vinson2008,ellard-gray2015}. We communicated the purpose of the study to all participants, ensuring they understood what their involvement entailed. 
Prior to the survey, we provided a disclaimer regarding the topics covered and included a trigger warning for participants who might be sensitive to discussions of negative experiences related to their sexuality. 
To prioritize participant comfort and well-being, all questions were optional, and participants were free to discontinue the survey at any point should they feel uncomfortable. 

\subsection{Threats to Validity}
Construct validity threats arise primarily from response bias, where participants provide inappropriate answers, compromizing the data integrity. While the study mitigates this risk by leveraging a relatively large sample of a hidden population, the inherent diversity of participants, such as institutional, geographical, and cultural factors, may still limit generalizability~\cite{baltes2021}.
Internal validity is at risk due to selection bias, a concern typical in studies involving hidden populations. The sampling methods, which rely on self-selection, may over-represent certain groups of individuals who are more open to share their experiences, thus, they may not fully represent the entire spectrum of queer individuals in software engineering and humanities. Response bias is another threat, as participants who choose to engage with the survey may systematically differ from those who do not. This selective engagement could influence the overall portrayal of queer student experiences, and should be confirmed with more research.
External validity threats primarily stem from social desirability bias, where participants might adjust their responses to align with perceived social norms or expectations, rather than providing genuine insights. To mitigate this, the survey was designed to ensure anonymity and provide a welcoming environment for respondents. However, it is important to acknowledge that these biases may still influence the findings and limit their generalizability to broader populations.
Since both authors independently completed the analysis and reviewed the coding for the thematic analysis, we strengthened the conclusion validity by reducing potential bias and ensuring consistency in interpreting the data. To enhance the reproducibility of our study, all non-sensitive materials are available for verification.\footnote{\url{https://figshare.com/s/dc18f812716249d58357}}




\section{Results}\label{sec:results}
%
%

\subsection{RQ1: Outness and Campus Climate}
\begin{figure}
	\centering
	\includegraphics[width=1\linewidth]{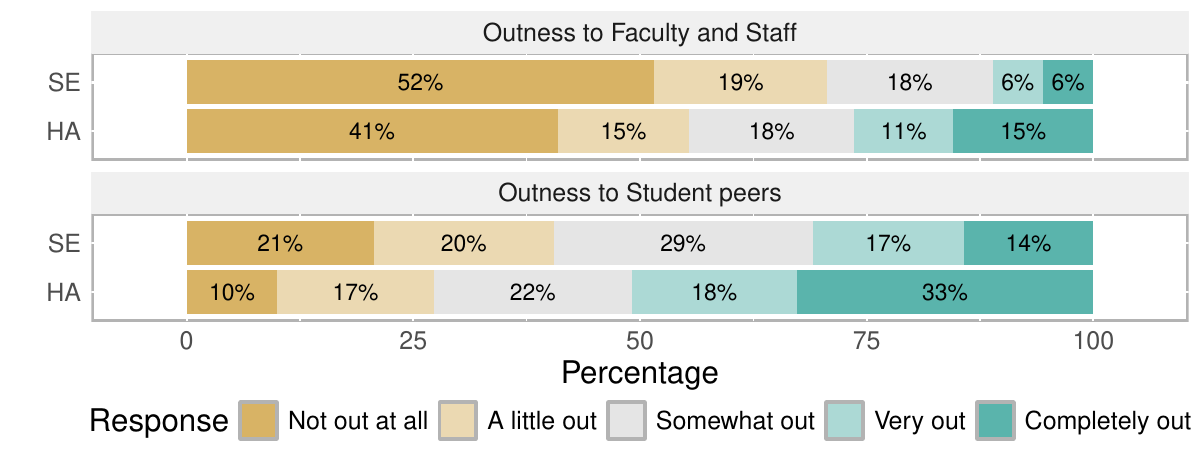}
	\caption{Level of \emph{outness} of queer students.}
	\label{fig:outness}
\end{figure}

\paragraph{\textbf{Level of Outness}}
\Cref{fig:outness} shows the level of outness of queer students.
In comparison to queer humanities students, queer software students are significantly less likely to be out with members of the faculty and staff ($p=0.01, \hat{A}_{12}=0.41$) as well as with their student peers ($p<0.001, \hat{A}_{12}= 0.37$). 

This may be due to a perception that being open about one's sexuality could impact their personal or academic relationships in a predominantly male-dominated field, leading to a reluctance to disclose private information. 
The humanities may foster a more accepting environment where students feel safer expressing their sexual orientation.

\begin{table}[t]
    \caption{The percentage reflects the proportion of queer students reporting a \emph{negative} campus climate (\emph{CC02}).}
    \label{tab:reasonsNegativeCampus}
    \begin{tabular}{p{4.2cm}rlrl}
        \toprule
    Category &
      \multicolumn{1}{l}{SE \%} &
       &
      \multicolumn{1}{l}{HA \%} &
       \\ \midrule
    Social isolation and exclusion &
      12 &
      \DrawPercentageBar{0.12} &
      3.63 &
      \DrawPercentageBar{0.363} \\
    Hidden identity 
    &
      20 &
      \DrawPercentageBar{0.20} &
      10 &
      \DrawPercentageBar{0.10} \\
    General sexism &
      5.6 &
      \DrawPercentageBar{0.056} &
      0 &
      \DrawPercentageBar{0} \\
    Bullying/harassment/mockery &
      9.6 &
      \DrawPercentageBar{0.096} &
      10.90 &
      \DrawPercentageBar{0.0109} \\
    Passing as non-queer &
      1.6 &
      \DrawPercentageBar{0.016} &
      0.90 &
      \DrawPercentageBar{0.090} \\
    Supportive inclusive Environments &
      9.6 &
      \DrawPercentageBar{0.096} &
      24.54 &
      \DrawPercentageBar{0.2454} \\
     Direct impact on Academic Context &
      5.6 &
      \DrawPercentageBar{0.056} &
      1.81 &
      \DrawPercentageBar{0.0181}\\
      \bottomrule
    \end{tabular}
    \end{table}

    \begin{table}[]
        \caption{The percentage reflects the proportion of queer students reporting a \emph{positive} campus climate (\emph{CC04}).}
        \label{tab:reasonsPositiveCampus}
        \begin{tabular}{p{4.2cm}rlrl}
        \toprule
        Category                                    & \multicolumn{1}{l}{SE \%} &                           & \multicolumn{1}{l}{HA \%} &                            \\ \midrule
        Social acceptance and respect   & 8                         & \DrawPercentageBar{0.08}  & 28.18                     & \DrawPercentageBar{0.2818} \\
        Inclusive lab environments & 7.2                       & \DrawPercentageBar{0.072} & 14.55                     & \DrawPercentageBar{0.1455} \\ 
        Inclusive initiatives and engagement                 & 2.4  & \DrawPercentageBar{0.024} & 9.09  & \DrawPercentageBar{0.0909}  \\ 
        Interactions and encouragement  & 37.6 & \DrawPercentageBar{0.376} & 31,82 & \DrawPercentageBar{0.3182} \\ 
        Positive recognition           & 5.6                       & \DrawPercentageBar{0.056} & 2.73                      & \DrawPercentageBar{0.0273} \\
     Neutral environment       & 10.4                      & \DrawPercentageBar{0.14}  & 10.91                     & \DrawPercentageBar{0.1091} \\
        Hidden identity                             & 11.2                      & \DrawPercentageBar{0.112} & 5.45                      & \DrawPercentageBar{0.0545} \\
        Limited interaction                         & 5.6                       & \DrawPercentageBar{0.056} & 0.91                      & \DrawPercentageBar{0.2454} \\
        No positive at all                          & 2.4                       & \DrawPercentageBar{0.24}  & 0                         & \DrawPercentageBar{0}     \\
        \bottomrule
        \end{tabular}
        \end{table}
The campus climate reflects the overall atmosphere and experiences that students encounter within their educational environment, reflecting both the social dynamics and the inclusivity within the institutions.

\paragraph{\textbf{Negative Campus Climate}}
Overall, the majority of queer student groups did not experience negativity regarding the campus climate. However, despite queer software students being less open about their sexuality at university, they report levels of exclusion and disadvantages that are quite similar to their peers in the humanities (HA: 17.46\%, SE: 15.45\%). In contrast, only 4.65\% of non-queer students (SE: 5.41\%, HA: 0\%) reported having faced negativity due to their sexuality. Neither the chi-square ($p=0.81$) nor the logistic regression revealed significant differences, likely due to the comparable levels of negativity experienced across groups.

The underlying reasons for the negative experiences reveal distinctions between the two queer groups (\Cref{tab:reasonsNegativeCampus}).


%
Notably, 20\% of software students, compared to 10\% of humanities students, attribute their negative experiences to hidden identities (\Cref{tab:reasonsNegativeCampus}), highlighting the complexities of navigating sexual orientation in a field that traditionally emphasizes technical skills over personal identity. 

In addition to concealing their authentic selves, software students suffer from social isolation (12\%) compared to the humanities students (3.63\%, \Cref{tab:reasonsNegativeCampus}). This is highlighted by one software student: \emph{“I have faced negative and exclusionary behavior, including being ignored or shunned by peers and faculty, which has hindered my ability to concentrate on my studies and affected my well being." (P767, SE)}.

Despite being less out, software students experience bullying, harassment, and discrimination at similar rates as humanities students (SE: 9.6\%, HA: 10.90\%). This indicates that both fields face similar barriers when it comes to acceptance. Examples of this discrimination include:
\emph{“I got called names in a group discussion and my point was not welcomed due to my sexual identity." (P848, HA)}.
\emph{“When I was about to present my presentation about gender performativity, one of the conservative studends complained to another "again about THOSE GAYS". (P467, HA)}.
Similar experiences were shared by the software students:
\emph{“Some professors would make jokes about sexualities." (P757, SE)}.
\emph{“I got a LOT of sexual comments, especially at first. And comments or people insinuating I only got where I am due to me filling "quotas" or that my technical skills aren't that good etc..." (P553, SE)}.
The prior factors highlight a distinct challenge encountered by queer software engineering students is the intersection of sexism~\cite{sczesny2006} with their sexual identity. One example of the 5.6\% of software students who face sexism show similar experiences as in industry~\cite{vanbreukelen2023a}: \emph{“No one was in any way insulting or condescending towards my sexual orientation or identity. It's only as a woman that people look at you stupidly: As a she/her, you get condescending looks from male fellow students or even male lecturers, and I've also heard stupid comments from some male students, such as ‘no wonder she doesn't understand' or ‘the course is so easy, why do others study so much for it' (with a view to the female fellow students)." (P286, SE)}. 
This gender bias extended to interactions with professors, as another software student recounted: \emph{“A professor made a comment about women need something 'softer and more fun' than computer science." (P561, SE)}.

Although we explicitly asked students about negative experiences many were unable to identify any and responded with positive reflections instead. Notably, 24.55\% of humanities students reported benefiting from a supportive and inclusive environment, compared to only 9.6\% of software students. 
This disparity indicates that while both groups face obstacles, humanities students may have access to a stronger network of support, possibly due to the emphasis on community-building and networking in their academic discipline as one humanities' student explained: \emph{“In my field of study, it seems to be a lot of non-heteronormative students which contributes to people being more accepting and sympathetic." (P529, HA)}.


\paragraph{\textbf{Positive Campus Climate}}
Queer students across both fields of study reported similar rates of positivity towards their sexuality, as reflected in the chi-square test, which revealed no significant difference between disciplines~($p=0.5$, SE: 60.32\%, HA: 65.45\%). 
However, when comparing the level of outness (\emph{SE01--02}) with the presence of positive experiences, we observed a significant difference through logistic regression analysis. Specifically, greater disclosure of sexual orientation to peers was associated with increasingly positive experiences among students ($p<0.001, \beta=0.42$), independent of their field of study ($p=0.94, \beta=-0.02$).

Despite this overall positivity, the underlying factors contributing to a positive campus climate differ~(\Cref{tab:reasonsPositiveCampus}). 
Notably, 37.6\% of software students and 31.82\% humanities students highlight positive interactions and encouragement from peers (\Cref{tab:reasonsPositiveCampus}), indicating a sense of collaboration exists across disciplines, particularly in software engineering, where teamwork is integral to daily work. As one software engineering student reflected: \emph{“I am lucky enough to become part of a queer group - where there were other asexual people - as well as other sexualities. The other people in the group were also very accepting and we could openly discuss things without feeling weird or rude - being able to ask each other questions about their sexuality as well as express our own troubles." (P787, SE)}. Another student shared a similar sentiment, noting: \emph{“Founding that other peers were part of the collective and having no fest of outing ourselfs was realiving." (P523, SE)}

In contrast, humanities students placed more emphasis on social acceptance and respect from peers and faculty, with 28.18\% noting a welcoming environment (\Cref{tab:reasonsPositiveCampus}) where one is \emph{“[...] being welcomed and acknowledged by peers and faculty, which has significantly boosted my learning experience and overall well being." (P767, SE)}.

This stands in contrast to the 8\% software students, suggesting that humanities disciplines may cultivate a stronger atmosphere of inclusivity and acceptance. This was highlighted by one software student: \emph{“My field is very technical so I don't know how it can make me feel welcomed because we don't really put emotion into the studying." (P29, SE)}.

Engagement in inclusivity initiatives, such as events promoting visibility for diverse identities, also differed between disciplines. Humanities students were more likely to resonate with these initiatives (9.09\%, SE: 2.4\%), as illustrated by the following comment: \emph{“Some professors openly wearing pins with the gay flag during pride month." (P496, HA)}.
This proactive approach in humanities programs appears to create more opportunities for students to connect, engage, and express their identities openly, contributing to a more inclusive campus culture. One humanities' student noted: 
\emph{“The inclusive language that is used on a regular basis and also seminars about intersectionality and queer culture." (P279, HA)}.

Inclusivity was also more evident in the labs and workspaces of humanities students (14.55\%), compared to the lower support reported by software students (7.2\%, \Cref{tab:reasonsPositiveCampus}). A humanities' student shared their experience in a translation class:
\emph{“In my translation class, I've had teachers point out that if we're translating and the gender isn't clear of who the partner is, then we have to be careful if we're translating into a gendered language because the default isn't male-female partnerships and we need to get as much information as possible about the relationship to translate it properly." (P828, HA).}
%
%
The presence of supportive professors was another major theme among humanities students:
\emph{“Multiple proffesors showed a welcoming attitude towards LGBTQ+ community and mentioned gay and lesbian writers and artists during their lectures in a positive light. I once met a proffesor at a protest event after a shooting of two LGBTQ+ people in my country." (P504, HA)}.
\emph{“We had a "welcome" game where lgbtq+ topics were talked about and we could easily put our experience in, no one was shocked or reacted negatively, even showed support. Also we have many queer professors." (P846, HA)}.

This disparity points to a potentially less collaborative atmosphere in software programs, where technical skills may overshadow social dynamics. 
However, positive interactions and inclusivity still made a difference, as evidenced by one software student: \emph{“I have also been in class with another instructor that really encouraged respect and inclusivity. This class happens to be one of those that i ended up performing best." (P515, SE)}.




In summary, the positive campus climate is influenced by various factors, including social acceptance, supportive environments, and positive interactions. The overall perception of inclusivity and acceptance appears to be stronger within humanities disciplines, revealing opportunities for software programs to enhance their supportive structures and foster a more welcoming atmosphere for all students.

\rqsummary{RQ1}{Queer software students are less out compared to humanities students, but experience a similar amount of negativity due to their sexuality. The more out students are, the more positive experiences they encounter.}

%
\subsection{RQ2: Sense of Belonging}
\begin{figure}
	\centering
	\includegraphics[width=1\linewidth]{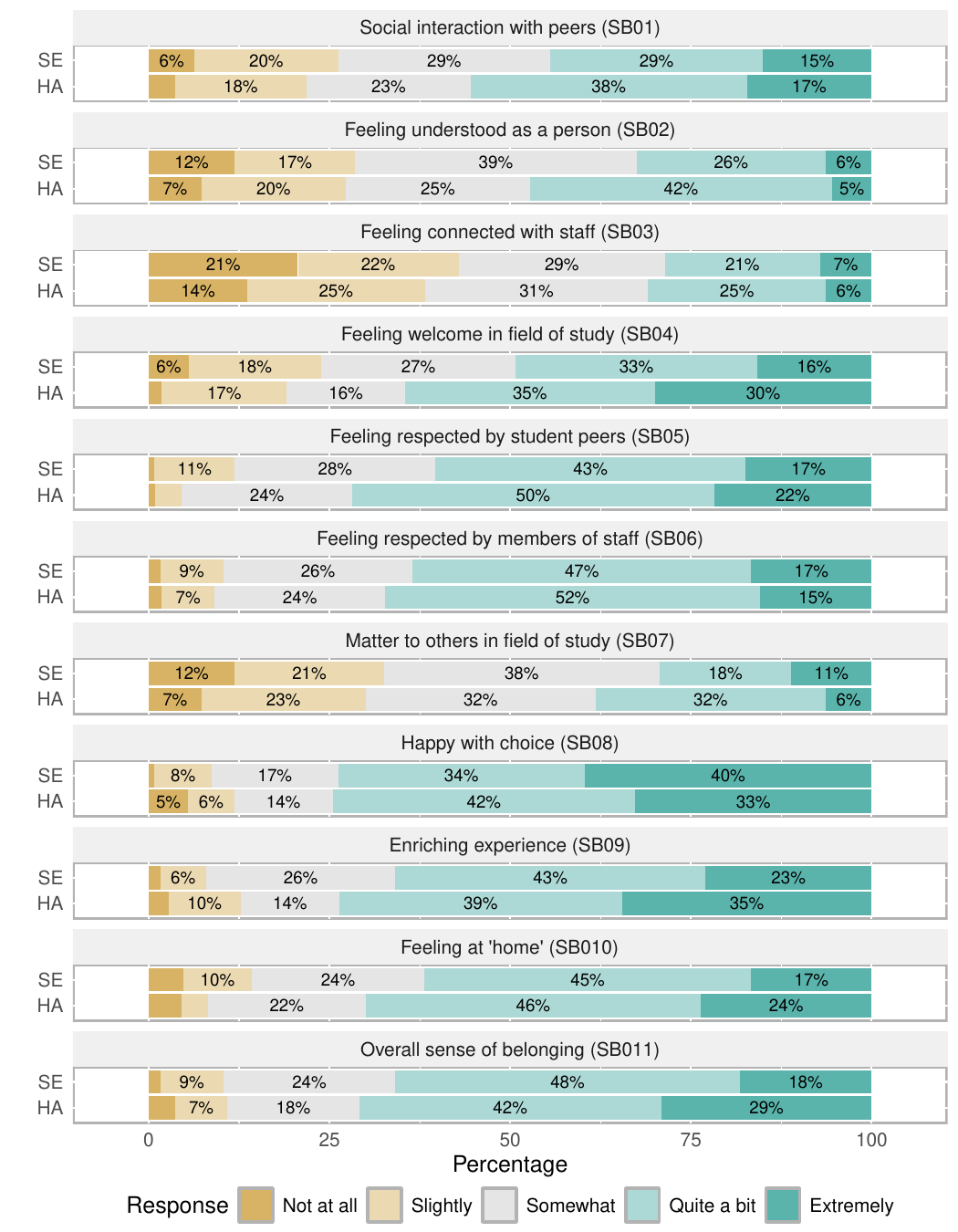}
	\caption{Comparison of \emph{sense of belonging} between queer software (SE) and humanities (HA) students.}
	\label{fig:sb-queer}
\end{figure}

\begin{table}[tb]
	\centering
	\caption{Impact of discipline and level of \emph{outness} of queer students on \emph{sense of belonging} (\Cref{tab:questionnairestudents}).}
	\label{tab:sb_outness}
	\resizebox{\columnwidth}{!}{%
		\begin{tabular}{lrrrrrr}
			\toprule
			& \multicolumn{2}{c}{Discipline} & \multicolumn{2}{c}{Out to Staff (\emph{SE01})} & \multicolumn{2}{c}{Out to Peers (\emph{SE02})} \\
			& $p$ & $\beta$ & $p$ & $\beta$ & $p$ & $\beta$ \\
			\midrule
			\emph{SB01} & 0.95 & -0.01 & 0.69 & 0.03 & \textbf{$<0.01$} & 0.32 \\
			\emph{SB02} & 0.96 & -0.01 & 0.25 & 0.07 &  \textbf{$<0.01$} & 0.27 \\
			\emph{SB03} & 0.70 & -0.06 & \textbf{0.01} & 0.19 & \textbf{0.04} & 0.14 \\
			\emph{SB04} & 0.15 & 0.21 & 0.06 & 0.13 & \textbf{0.01} & 0.17 \\
			\emph{SB05} & 0.23 & 0.14 & 0.14 & 0.08 & 0.13 & 0.08 \\
			\emph{SB06} & 0.67 & -0.05 & \textbf{$<0.01$} & 0.16 & 0.90 & 0.01 \\
			\emph{SB07} & 0.45 & -0.10 & \textbf{0.05} & 0.13 & \textbf{$<0.01$} & 0.24 \\
			\emph{SB08} & 0.12 & -0.22 & 0.20 & 0.08 & 0.40 & 0.05 \\
			\emph{SB09} & 0.76 & 0.04 & 0.19 & 0.08 & 0.18 & 0.08 \\
			\emph{SB10} & 0.56 & 0.08 & 0.12 & 0.10 & \textbf{0.02} & 0.14 \\
			\emph{SB11} & 0.96 & 0.01 & 0.47 & 0.04 & \textbf{0.01} & 0.17 \\
			\bottomrule
		\end{tabular}%
	}
\end{table}

To investigate the sense of belonging among software and humanities students, we analysed responses from our questionnaire (\Cref{tab:questionnairestudents}, \emph{SB01--SB11}).

As illustrated in \Cref{fig:sb-queer}, queer software students consistently reported lower scores across all aspects of their sense of belonging compared to their queer peers in the humanities. 
In particular, queer software students feel significantly less welcome (\textit{SB04}, $p=0.01,$ Â$_{12}$$=0.4$) and respected by their student peers (\textit{SB05}, $p=0.05$, Â$_{12}$ = 0.43) than the humanities students (\Cref{fig:sb-queer}). 
To further understand these dynamics, we conducted a sexuality x studies ANOVA. Although we did not identify a significant interaction with sexuality and studies, the results show a significant impact of a students queer sexuality (\textit{SB05}, $p = 0.02$, $F(1, 1) = 5.18$) as well as their studies being software-related (\textit{SB05}, $p = 0.05$, $F(1, 1) = 4.08$) on how much respect they felt from their peers. A students queer sexuality  effects the amount of respect they felt from members of staff in their current field (\textit{SB06}, $p = 0.01$, $F(1, 1) = 6.13$).

The analysis of students' level of outness reveals distinct effects on their sense of belonging, depending on the social context (\Cref{tab:sb_outness}). Greater outness towards faculty and staff (\textit{SE01}) is significantly associated with increased feelings of connection (\textit{SB03}, $p = 0.01, \beta = 0.19$) and respect (\textit{SB06}, $p < 0.001, \beta = 0.16$). In contrast, \Cref{tab:sb_outness} shows that outness towards peers (\textit{SE02}) is linked to higher levels of peer interaction (\textit{SB01}, $p < 0.001, \beta = 0.32$), stronger perceptions of being understood (\textit{SB02}, $p < 0.001, \beta = 0.27$), and increased feelings of being welcomed (\textit{SB04}, $p = 0.01, \beta = 0.17$) and at home (\textit{SB10}, $p = 0.02, \beta = 0.14$). Students who are more out report a significantly stronger sense of belonging (\textit{SB11}, $p = 0.01, \beta = 0.17$, \Cref{tab:sb_outness}), independent of their studies.

\rqsummary{RQ2}{Queer software students experience a lower sense of belonging compared to their peers in humanities, particularly regarding their feelings of being welcomed and respected by fellow students. A higher level of outness, though, results in a higher sense of belonging for all students.}

%
\subsection{RQ3: Dropout Rate} 
%


\begin{figure}
    \centering
    \includegraphics[width=0.75\linewidth]{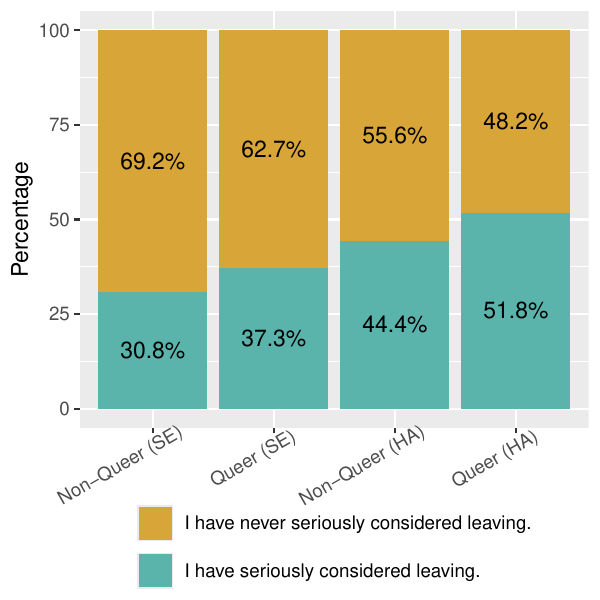}
    \caption{Students' thoughts about dropping out of their field.}
    \label{fig:dropout}
\end{figure}
A low sense of belonging is often associated with higher dropout rates~\cite{stout2016lesbian,schefer2023study}. To investigate this phenomenon, we analysed students’ thoughts on leaving their field of study~(\emph{DP01}, \Cref{tab:questionnairestudents}). 

As shown in \Cref{fig:dropout}, queer humanities students reported significantly more thoughts of leaving their field of study compared to the software students ($p=0.03$).
%
When examining the influence of sexuality and field of study on all students thoughts about dropping out, our logistic regression analysis revealed significant results ($p=0.02, \beta=-0.59$), indicating that humanities students think about leaving more often. 
In contrast, sexuality itself did not show a significant impact on these thoughts ($p=0.39, \beta=0.29$), however, it still results in a positive association, suggesting that queer students think about leaving more often than their non-queer peers. 
This disparity may be partly due the observation that students in software-related subjects who have thought about leaving, might have already left, since most students dropout in the first semester \cite{schefer2023study}, and in our data set only 7\% of computer science students are in their first semester.

Further logistic regression analysis incorporating the sense of belonging (\emph{SB01--11}) as a predictor revealed that particularly students who express dissatisfaction with their field of study demonstrate a significantly higher probability of considering dropping out (\emph{SB08}, $p=0.03, \beta=-0.32$). A similar trend was noted regarding students’ perceptions of how enriching their educational experience has been so far (\emph{SB09},$ p=0.07, \beta=-0.27$).


The reasons behind queer students leaving their studies (\emph{DP02}, \Cref{tab:questionnairestudents}) reveal a complex combination of academic, social, and mental health challenges. 
Among queer software students, the primary reason for leaving, is the inherent difficulty of the subject matter, which they find overwhelming (31.91\%): \emph{“The courses were sometimes very difficult. many failed attempts. the feeling of being ‘too stupid'." (P286, SE)}. In addition to the perceived difficulty, the feeling of being not capable of succeeding correlates with the academic persistence~\cite{tao2019}. This highlights the need for a more supportive learning environment that considers students’ diverse backgrounds and abilities.

Almost one third (29.78\%) reported feelings of social isolation and a lack of belonging within their academic community, as one student mentions: \emph{“I don't feel supported by my faculty, it's quite isolating so some mental health problems arise and i felt like at some points it would be better for me to leave or take a break from my studies." (P794, SE)}. 
This issue goes along with mental health issues which also emerged as a significant concern for 25.53\% of queer software students: \emph{“Because I was seriously depressed at not having friends in my current university." (P500, SE)}. This emphasizes the importance of fostering connection and support networks~\cite{schefer2023study}. 

Queer humanities students identified mental health challenges as their leading reason for leaving (36.84\%). This concern intersects with the 35.08\% who reported a mismatch between their studies and personal interests or expectations, indicating that the curriculum may not adequately engage or resonate with their aspirations as one linguistic student describes: \emph{“Eventhough I love linguistics and languages I kinda felt like I don't want to pursue a career in that field, like maybe it wouldn't give me as profitable job as a different field. I also feel like the curriculum is overloaded with filler subjects in my course." (P813, HA).}

Interestingly, our analysis of the factors motivating queer students to remain in their studies revealed that a strong passion for their respective fields is a significant determinant. Specifically, 55.69\% of queer software students and 71.69\% of queer humanities students stated their enthusiasm for their studies as primary factor in their decision to persist.   

However, notable differences emerged between both groups regarding their motivations. Queer software students are often driven by a sense of determination~\cite{deci2000} and a long-term vision for their professional development, particularly in terms of acquiring specific skill sets necessary for success in the tech industry (25.31\%). 
Queer humanities students tend to prioritize external goals, such as achieving financial stability or meeting societal expectations (15.09\%). This contrast highlights how intrinsic motivations can differ across disciplines, showing how educational institutions might better support queer students in maintaining their academic journeys.

\rqsummary{RQ3}{Queer humanities students think about dropping their studies more than their software peers. Both queer student groups suffer from mental health issues and dissatisfaction with their field of study.}

\section{Discussion}\label{sec:discussion}
A key theme that emerged across all research questions is the need to \emph{create a welcoming and safe environment for queer software students}. The strategies below provide a starting point for emphasizing the human aspect of these studies.

\subsection{Enhance Visibility and Recognition of Queerness}
Visibility is key to establish a welcoming environment~\cite{gonzalez2024,de2023lgbtqia}! If issues of sexual identity and the challenges surrounding them are ignored, the \emph{diversity crisis} will persist, and continue in industry~\cite{desouzasantos2023}, as these topics are often overshadowed by heteronormative assumptions~\cite{miller2021,leyva2022}.
Promoting queer events, such as \emph{Pride Month}, and integrating these celebrations into the academic curriculum are effective ways to increase visibility. As some humanities students pointed out, even small gestures---such as displaying rainbow flags or wearing a pin---can make a meaningful difference (\emph{RQ1})~\cite{gonzalez2024}.

All university stakeholders should work more collaboratively~\cite{miller2021} to make students feel they matter (\emph{RQ2}). The support of leadership figures and their belief in students (\emph{RQ1}) can improve students' confidence~\cite{cech2009} and help mitigate feelings of imposter phenomenon due to heteronormative ideals~(\emph{RQ1})~\cite{tao2019,guenes2024}.
Simple measures, such as recognizing queer figures and contributions in lectures, can raise awareness with minimal effort. Just as humanities students reported hearing about \emph{“lesbian writers and artists during lectures in a positive light”} (\emph{RQ1}, P504, HA), software engineering courses could highlight for instance Alan Turing or Tim Cook. 

Additionally, using inclusive language in the classroom, e.g., using the phrasing \emph{partners} in modelling samples or discussion rather than assuming heteronormative structures (\emph{RQ1})---\emph{“It feels nice when professors/other staff give examples and/or some rules or directly speak to you and mention something like "and your boyfriend or girlfriend ". It doesn't make a big change but makes you feel more welcomed. (P816, SE)}.

\subsection{Beyond Tech: Foster a Social  Community}
Building a supportive campus climate is essential for cultivating a sense of community among queer students~\cite{baumeister2017}.

Establishing support structures such as mentoring programs~\cite{sulimani-aidan2024,sarna2021} or queer student organisations can enhance social cohesion and reduce the lack of belonging (\emph{RQ2}). The data reveals that many students, particularly in software engineering, struggle to form friendships in the technical environment (\emph{RQ1}). The importance of social connections was underscored by students who noted that having supportive peers helped them navigate their challenges (\emph{RQ3}).
For instance, professors can mention support groups or resources at the beginning of lectures, making students aware that there are safe spaces available if they feel uncomfortable (\emph{RQ1}). Additionally, one could introduce icebreaker activities into team work assignments. By normalizing social interactions as part of the software development process~\cite{hughes2017}, students can foster connections with their peers before engaging in more technical tasks like coding or requirements analysis. This shift not only builds social ties but challenges stereotypes about software engineering being an impersonal (\emph{RQ3}, P799, SE), purely technical field~\cite{spieler2020f,wong2023}, which might be transferable to mitigating feelings of isolation in professional software development~\cite{desouzasantos2023}.
Additionally, the discussion of socio-technical issues in tech classes should be more intentional in order to highlight the interplay of understanding culture, identity, and empathy in  tech-driven environments.

\subsection{Prioritize Mental Health and Emotional Well-Being}
Mental health is a critical area that requires more attention in educational institutions and academia~\cite{penzenstadler2024well}. Our findings indicate that queer humanities students are more likely to discuss mental health openly (\emph{RQ1}), likely influenced by higher female representation and a cultural tendency toward emotional expression~\cite{stewart2021,roseth2019features}. In contrast, the male-dominated software engineering environment may contribute to a reluctance to openly address mental health concerns~\cite{kosciw2015}.

Encouraging open discussions about mental health, \emph{normalizing these conversations}, and providing accessible resources can foster a culture of understanding and acceptance~\cite{deci2000,penzenstadler2024well}.
Again, communication and empathy are key. Even if some faculty members do not feel personally connected to these issues, it is essential to acknowledge and highlight their importance to others. 
%
In return, humanities students can learn from software students’ a greater determination by adopting a more goal-oriented mindset. Software programs often connect studies to concrete, real-world outcomes and clear career paths, which might build more resilience. Humanities could do the same by offering more projects or interdisciplinary work, helping students persist through challenges.


\section{Conclusions and Future Work}\label{sec:conclusions}

This study aimed to gather insights into how the experiences of queer software engineering students could be improved by comparing them with those of queer students in humanities disciplines. Our findings reveal that queer software engineering students report a significantly lower \emph{sense of belonging}, particularly in terms of feeling welcomed and respected by their peers. Additionally, they are less likely to be open about their sexual identity with both faculty and fellow students. In contrast, queer students in the humanities, while reporting more positive experiences exhibit a higher dropout rate compared to their software counterparts who showed more resilience to stay in their studies.
Our data suggest that a lack of visibility and institutional support exacerbates the marginalization of queer students. This highlights the critical need for universities to foster environments that not only embrace diversity but actively support and \emph{celebrate} queerness.

%
Further research should examine how queer students develop \emph{coping} and \emph{passing strategies} in response to negative experiences, and whether those differ from queer professionals in industry. These strategies, in turn, can impact their well-being and ability to manage stress---an especially relevant consideration in software development~\cite{grassl2023exposing}. In this context, we should also explore how experiences vary based on course content, teaching approaches, and teamwork~\cite{grassl2023diversity}.
Since hardships can sometimes lead to \emph{self-empowerment}~\cite{yarberry2021impact}, further studies should explore the \emph{determination} exhibited by queer software students, and how those interact with intersectionality~\cite{vanbreukelen2023a,leyva2022}.
We should also explore queer identities not just as a whole, but focus on the unique experiences tied to specific sexual identities~\cite{santos2024diversity}. Additionally, exploring variations in appearance and behaviour that challenge heteronormative standards~\cite{sczesny2006}, along with cultural and geographical differences, can provide a more comprehensive understanding.
Finally, this research would also benefit from qualitative insights through interviews with queer students who participated in this study’s survey. Many respondents expressed a willingness to support this kind of research and share their experiences more deeply.
By continuing this line of research, we take an important step toward creating more inclusive learning environments where queer students must not hide their identities, but instead feel empowered to be their authentic selves. Ultimately, sexuality should not dictate how students navigate their academic careers---it should indeed be irrelevant to their opportunities for success and inclusion.

\section*{Acknowledgment}
We thank all our participants for being open and sharing their valuable experiences. Their resilience and determination to keep moving forward inspire positive change!

\bibliographystyle{IEEEtran}
\bibliography{related}

\end{document}